\documentclass[twocolumn,showpacs,preprintnumbers,amsmath,amssymb]{revtex4}

\usepackage{graphicx}
\usepackage{dcolumn}
\usepackage{bm}
\usepackage{pifont}

\begin{document}

\preprint{APS/123-QED}

\title{Star-Like Micelles with Star-Like Interactions:\\
 A quantitative Evaluation of Structure Factors and Phase Diagram}

\author{M. Laurati}
\author{J. Stellbrink}
\author{R. Lund}
\author{L. Willner}
\author{D. Richter}%
\affiliation{%
Institut f\"ur Festk\"orperforschung, Forschungszentrum J\"ulich\\
52425 J\"ulich, Germany}%
\author{E. Zaccarelli}
\affiliation{Dipartimento di Fisica and  INFM-CRS SOFT, Universit\`a di Roma La Sapienza, P.zza A. Moro 2, I-00185, Roma, Italy}%
\date{\today}

\begin{abstract}
PEP-PEO block copolymer micelles offer the possibility to investigate phase behaviour and
interactions of star polymers (ultra-soft colloids). 
A star-like architecture is achieved by an extremely asymmetric block ratio (1:20). 
Micellar functionality $f$ can be smoothly varied by changing solvent composition (interfacial tension). 
Structure factors obtained by SANS can be 
quantitatively described in terms of an effective potential developed for star polymers. 
The experimental phase diagram reproduces to a high level of 
accuracy the predicted liquid/solid transition. Whereas for intermediate  $f$ a bcc phase is observed, 
for high $f$ the formation of a fcc phase is preempted by glass formation.
\end{abstract}

\maketitle

Star polymers, \textit{i.e.} $f$ polymer chains tethered to a central microscopic core, 
can be regarded as ultra-soft colloids 
bridging the properties of linear polymer chains and
hard sphere colloids \cite{grest:acp:96}.  
The hybrid character is reflected in their 
effective interaction potential introduced by 
Likos et al. \cite{likos:prl:98}, 
\begin{eqnarray}
\label{eq-potential}
\beta V(r) & = & {5\over 18}f^{3/2}\left[ -\ln\left({r 
\over\sigma}\right) + {1 \over {1+\sqrt f/2}}\right],~~~~~~~~r \le \sigma \nonumber\\
 & = & {{5} \over 18}f^{3/2} {{\sigma / r} \over {1+\sqrt 
f/2}}\exp\left[-{{\sqrt f (r-\sigma)}\over{2\sigma}}\right],~~r \ge \sigma 
\nonumber\\
\end{eqnarray}
with $\beta=1/k_BT$, $r$ the distance between star centers 
and $\sigma$ the corona diameter. 
Eq. \ref{eq-potential} has 
given good results in modelling small angle neutron scattering (SANS) data of 
 star polymer solutions \cite{likos:prl:98, stellbrink:pcps:00}. 
The corresponding theoretical phase diagram \cite{wat:prl:99} shows unique features, 
such as different crystalline phases depending sensitively on  volume fraction $\phi$ and $f$. 
Recently the phase diagram has been revisited to include the glass transition 
as a result of dynamical arrest of the system \cite{emanu:prl:03}, 
corroborating several experimental studies \cite{stellbrink:pcps:00,vlass:jpcm:01,kapnistos2001}. 

Unfortunately, the high interest in star polymers is not reflected in their availability. 
Synthesis requires considerable preparative efforts, 
therefore it is of particular interest to substitute star polymers by an 
easy to establish model system. 
Ideally, this analogue should not only show the typical molecular architecture of star polymers, 
but also interact via the effective potential given in Eq. \ref{eq-potential}.

A few attempts to connect micellar systems to star polymers have been done, 
however, approaching the star-like regime \cite{hamley:1998,halperin:aps:1992} 
defined by the block ratio $N_{a}/N_{b}\gg 1$ 
(with a/b the soluble/insoluble block) is not trivial. 
Gast et al. investigated poly(styrene)-poly(isoprene) (PS-PI) block copolymers \cite{gast:prl:93}. 
They found experimental evidence for
fcc and bcc phase formation, 
with crystalline geometry determined by block composition ( $0.8 \le N_{a} / N_{b} \le 4$).  
Recently PS-PI \cite{lodge:prl:04} and
PEO-PBO \cite{hamley:cps:98} micelles showed a
temperature induced fcc-bcc transition. 
This transition could be related to changes in $f$ 
and compared to the phase diagram of star polymers 
 indicating some similarities in the interactions  
although the block ratio was rather small, $N_{a} / N_{b} \approx 2$ \cite{lodge:prl:04}.
Micelles formed by hydrophobically modified PEO with $N_{a} / N_{b} \approx 150$ have been investigated \cite{beaudoin02}, 
but interactions were characterized by mapping onto an equivalent hard sphere system. 
Therefore, a systematic study to 
correlate  the starlike architecture of the individual micelle  
with star-like interactions has never been performed.

In this Letter we show that structure factors of 
poly(ethylene-alt-propylene)-poly(ethylene oxide) (PEP-PEO) micelles 
(which fulfill  all prerequisites of the star-like regime \cite{willner:epl:00, reidar:macro, stellbrink:jpcm}) 
 provide the basis to investigate such a correlation.
SANS measurements at core contrast allow a direct determination of 
experimental structure factors $S(Q)$ which  can be described  starting from Eq. \ref{eq-potential} 
with parameters $\phi, f$, and $\sigma$ directly given by experimental values. 
In addition, we can reproduce to a high level of accuracy the 
liquid-solid transition of the theoretical phase diagram. 
This was done over a wide range of $f$ and spanning the 
range from dilute to concentrated solutions,  
providing a comprehensive characterization of block
copolymer micelles in terms of the microscopic, effective potential originally developed for star polymers. 

The micellization behavior of PEP-PEO in aqueous solution is governed by the high interfacial tension, $\gamma$=46 $m$N/m, 
between PEP and water and shows the following features \cite{poppe:macro:97,willner:epl:00}: 
i.) PEP-PEO forms micelles with the hydrophobic PEP constituting the core 
and the hydrophilic PEO constituting the corona. 
ii.) Micelles are even formed in a very asymmetric block composition with high PEO content ($N_{a} / N_{b} \approx 20$). 
iii.) All micellar cores
are completely segregated, \textit{i.e.} not swollen by  solvent. 
iv.) The micelles are kinetically frozen: Although chemically not linked,
exchange of block copolymers between different
micelles could not be observed, even not at elevated temperatures \cite{reidar:thesis}. 
This means that no free chains,  \textit{i.e.} no depletion effects \cite{stiakakis:prl:02} are present  
in contrary to ref. \cite{lodge:prl:04}. 
v.) Adjusting $\gamma$ by addition of a PEO selective cosolvent like 
N,N-dimethylformamide (DMF), allows to smoothly vary $f$ even in the star-like regime 
(where $f\sim\gamma^{6/5}$ was confirmed by experiments \cite{reidar:macro}). 

The asymmetric PEP-PEO block copolymer under study was synthesized by anionic
polymerization \cite{poppe:macro:97}. 
To exploit h/d-contrast variation in our SANS experiments, 
the individual blocks have been selectively protonated/deuterated, see Tab. \ref{tab:characterization}. 
\begin{table}
\begin{center}
\begin{tabular}{c c c c c c}
\hline
& $M_n$ & $M_W/M_n\footnote{overall polydispersity by SEC}$ & $D_p$ & $x_h$ & $\rho[10^{10}cm^{-2}]$\\ 
\hline
PEP (block) & $1100$ & $1.06$ & $15$ & $1.0$ & $-0.31$\\
PEO (block) & $20700$ & $1.04$ & $436$ & $0.11$ & $6.32$\\ 
\hline
\end{tabular}
\begin{tabular}{c c c c c c}
$x_{DMF}$ & $\rho_0[10^{10}cm^{-2}]$ & $f^{\,\,\,b}$ & $R_{g,core}^{\,\,\,b}$[\AA] & $R_g$\footnote{From form factor analysis}[\AA] & $R_g$\footnote{From $\bar{\sigma}$ values obtained from fits}[\AA]\\
\hline
 $0.0$ & $6.33$ & $136$ & $31$ & $194$ & $192$\\
 $0.1$ & $6.32$ & $94$ & $29.5$ & $179$ & $161$ \\
 $0.2$ & $6.34$ & $82$ & $29$ & $164$ & $154$\\
 $0.3$ & $6.33$ & $73$ & $27$ & $150$ & $140$\\
 $0.4$ & $6.31$ & $67$ & $27$ & $140$ & $136$\\
 $0.5$ & $6.31$ & $63$ & $26$ & $135$ & $127$\\
\hline
\end{tabular}
\caption{\label{tab:characterization}Characterization of PEP-PEO block copolymer and micelles.
 $D_p$ is the degree of polymerization, $x_h$ the protonation fraction and $\rho$ and $\rho_0$ the scattering length densities 
 of polymer blocks and solvent. 
 $x_{DMF}$ is the molar fraction of DMF in the solvent, $R_{g,core}$ and $R_g$ the core and overall micellar radius of gyration.}
\end{center}
\end{table}
Using water/DMF mixtures we prepared samples with
six different $f$ going from dilute to the very concentrated regime. 
Since the micelles are kinetically frozen, all samples were directly prepared in the corresponding solvent mixture 
and annealed for at least 1 week, 
for experimental details see \cite{reidar:macro}. 
SANS data were corrected following standard procedures and normalized to 
absolute units to allow a quantitative theoretical interpretation. 
\begin{figure}
%
%
\vspace{0.5cm}
\includegraphics[angle=270,scale=0.34]{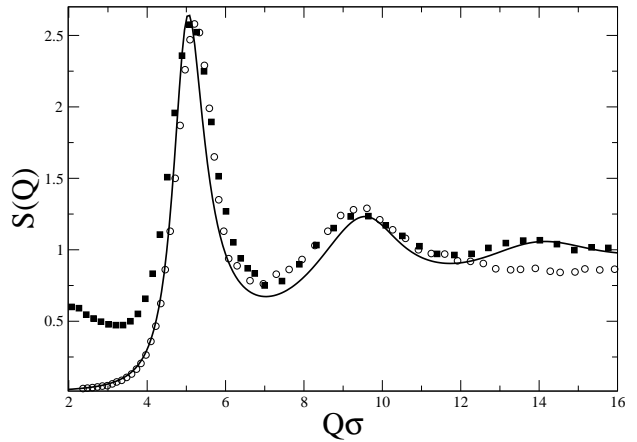}
\vspace{0.5cm}
\caption{\label{fig:sfcomparison} S(Q) for PEP-PEO micelles with 
$f=63$ ($\blacksquare$) compared to 64-arm (nominal) PB star ($\circ$), 
both samples have the same ratio $\phi/\phi^*\approx 1$. Solid line: Theoretical 
S(Q) calculated by applying the RY-closure for the OZ-equation starting from Eq. \ref{eq-potential}, see text.}
\end{figure}

The macroscopic scattering cross section, $({d\Sigma}/{d\Omega})(Q)$, measured by SANS can 
be expressed as a product of single particle contributions, the 
particle form factor $P(Q)$, and the structure factor $S(Q)$, which contains all 
information about particle interactions:
\begin{eqnarray}
\label{int-pq-sq}
\bigg(\frac{d\Sigma}{d\Omega}\bigg)(Q) = 
N_z\;P(Q)\;S(Q)
\end{eqnarray}
Here $N_z$ is the number density of particles. 
Experimental $S(Q)$ can be extracted from SANS data 
by dividing out the experimental $P(Q)$ measured in dilute solution.
However, this procedure is only valid, if  particle size and shape, 
\textit{i.e.} $P(Q)$, are unaffected by concentration. 
This is in general not the case for deformable particles like micelles, but we can 
overcome this problem by proper application of contrast variation techniques. 
Performing all SANS experiments in core contrast, \textit{i.e.} 
adjusting the scattering length density $\rho_0$ of the solvent by use of h/d-isotopic mixtures to that of PEO, 
reduces the contrast factor between corona and solvent to zero, see Tab. \ref{tab:characterization}. 
Only the compact PEP core, which is completely unaffected 
by increasing concentration, is ``visible'' in the SANS experiment.

Micellar characteristics, \textit{i.e.} functionality $f$, core radius $R_c$ and overall micellar radius $R_m$ have been determined as 
described in \cite{reidar:macro,stellbrink:jpcm}. 
$R_g^b$ in Tab. \ref{tab:characterization} is calculated according to $R_g^b = \sqrt{5 / 11} R_m$ valid for star polymers, 
$R_{g,core}$ according to $R_{g,core}= \sqrt{3 / 5} R_c$ valid for solid spheres. 

In this Letter we will  focus  on the results obtained from concentrated solutions, 
where particle interactions are dominant.  
Fig. \ref{fig:sfcomparison} shows the comparison between $S(Q)$ obtained 
for micelles with $f=63$ (water/DMF mixture with molar fraction $x_{DMF}=0.5$) and a corresponding poly(butadiene) (PB) star polymer in 
methylcyclohexane with (nominal) 64 arms \cite{stellbrink:pcps:00}. 
Both samples have been measured at a volume fraction $\phi\approx \phi^*$, 
with $\phi^* =3/(4\pi R_m^3)\times (f M_w/\overline{d} N_A)$ being the overlap volume fraction and 
$\overline{d}$ the average density. 
All main features of the star $S(Q)$ with respect to peak positions and heights 
are perfectly reproduced. 
The only difference is found in the low $Q$ region. 
Here the micellar $S(Q)$ shows  some increase possibly
indicating the presence of large scale structures in the micellar solution.  
These clusters might arise either from i.) (inherent) mesoscopic heterogeneities as also found for purely repulsive 
star polymer solutions \cite{stellbrink97,kapnistos2001} 
or ii.) from weak attractive interactions due to the decreasing solvent quality with increasing $x_{DMF}$ \cite{reidar:macro}. 
At the moment the physical origin of  clustering is not clear, but for case i.) this would only further support the analogy between
PEP-PEO micelles and star polymers, and 
for case ii.) recent theoretical results have shown that weak attractive interactions do not substantially influence the features 
in the region of the main structure factor peak \cite{federica:jpcm:03}. 
As will be shown below on the basis of our analysis, the low-Q scattering do not have major relevance 
on the quantitative interpretation of S(Q) for $Q \ge  10^{-2}\AA^{-1}$. 

Fig. \ref{fig:sfcomparison} also shows a fit to the experimental data using Eq. \ref{eq-potential},
with the corona diameter $\sigma$ being the only adjustable parameter. 
The theoretical $S(Q)$ was calculated by applying 
the self-consistent Rogers-Young (RY) closure for the Ornstein-Zernike equation. 
This closure has already been proven to be quite 
accurate compared to simulations of the same model potential for relatively 
small $f$ \cite{watz:jpcm:98,federica:jpcm:03}. 
Different S(Q) have been calculated for a fine grid of functionalities 
around the experimental $f=63$ to take into account the error on $f$. 
The experimental uncertainty on $f$  is such that no significant 
changes in $S(Q)$ are observed. 
From this fit we found a corona diameter $\sigma = 226 \AA$. 
The radius of gyration $R_g$ can be calculated by 
$R_g = \sigma / 2+\sigma / \sqrt{f}$ \cite{likos:prl:98}. 
The obtained $R_g = 127\pm 7$ \AA\ from the fit is in very good agreement 
with $R_g = 135\pm 10$ \AA\ found from independent SANS measurements of 
the micellar form factor \cite{reidar:macro}. 
\begin{figure}
\includegraphics[angle=270,scale=0.36]{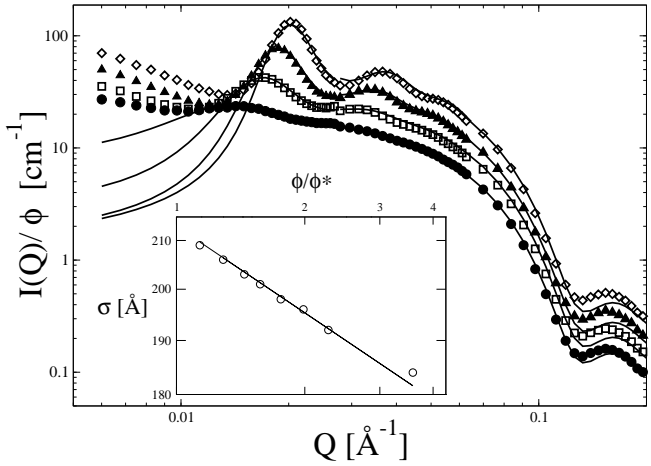}
\caption{\label{fig:int-fit} Experimental intensities I(Q) for $f=63$ at different $\phi$ in the fluid phase, $\phi<\phi^*$: 
($\bullet$) $\phi$=0.017, ($\square$) $\phi$=0.034, ($\blacktriangle$) 
$\phi$=0.051, ($\lozenge$) $\phi$=0.067 and fits (solid lines). 
For clarity I(Q) are divided by factors: ($\lozenge$) 1, ($\blacktriangle$) 1.5, ($\square$) 2, ($\bullet$) 3. 
Inset: volume fraction dependence of $\sigma$ for $f=63$, showing the expected scaling behavior $\sigma\sim\phi^{-1/8}$ for $\phi>\phi^*$.}
\end{figure}  
Data analysis was further refined by directly fitting $({d\Sigma}/{d\Omega})(Q)$ 
including the convolution to the 
experimental resolution function \cite{pedersen}. 
We assumed a compact core model for $P(Q)$, 
fixing the parameters to $f$ and $R_c$ values obtained from the characterization of the micelles. 
The validity of this assumption can be rationalized from Fig. \ref{fig:int-fit}, where the minimum 
arising from the solid core appears at $Q \approx 0.13 \AA^{-1}$ independent of $\phi$. 

On the basis of the good aggreement between theory and experiment we then
 applied this analysis procedure to all $f$ and $\phi$ studied. 
The excellent quality of the fits for $Q \ge  10^{-2}\AA^{-1}$ is shown in Fig. \ref{fig:int-fit} 
for samples of different $\phi$ but same functionality $f=63$. 
For $\phi\leq\phi^*$ we found a nearly constant corona diameter giving a
 mean value $\bar{\sigma}=226\pm 4$ , 
which corresponds to the low concentration, unperturbed corona diameter. 
$\bar{\sigma}$ was then used to transform the given experimental number density $N_z$ to the packing fraction $\eta = N_z \; \pi/6\; \sigma^3$. 
$R_g$ values calculated from $\bar{\sigma}$ are compared to results from form factor analysis in Tab. \ref{tab:characterization}.

\begin{figure}
\includegraphics[angle=270,scale=0.34]{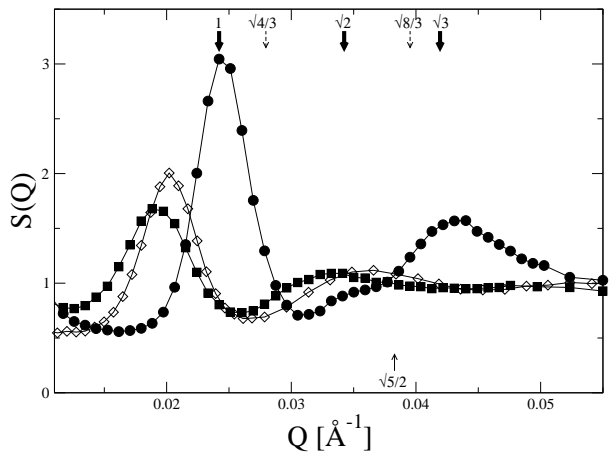}
\caption{\label{fig:sq-scaling} Experimental S(Q) for: ($\square$) $f=130$, ($\lozenge$) $f=82$ and ($\circ$) $f=67$ at $\eta\approx 0.22$, 
showing the crossover from gel to crystal with decreasing $f$. S(Q) for $f=67$ shows peak position ratios characteristic for bcc crystal (bold arrows). 
Expected peak positions for fcc (dashed arrows)  and A15 crystals. 
Solid lines: Guide to the eye.}
\end{figure}

Above $\phi^*$ we had to use a different approach for recalculating $\eta$. 
According to the Daoud-Cotton  theory \cite{daoud-cotton} the stars are expected to shrink as $\propto \phi^{-1/8}$, and indeed, 
the fit values for $\sigma(\phi\ge\phi^*)$ show exactly this scaling behaviour as shown in the inset of Fig. \ref{fig:int-fit}. 
Therefore we decided to use the individual $\sigma(\phi)$ for recalculating $\eta$. 
\begin{figure}
\includegraphics[angle=270,scale=0.34]{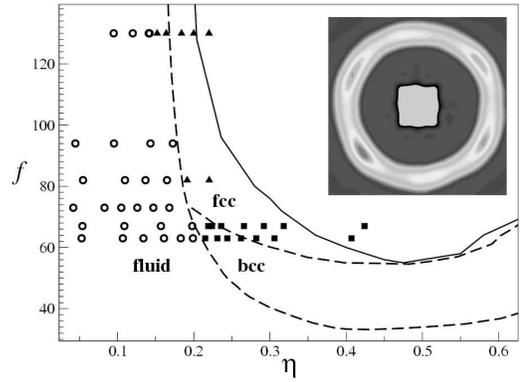}
\caption{\label{fig:phdiag} Experimental phase diagram (symbols) of star-like micelles  vs. theoretical 
phase diagram (lines) of star polymers. 
($\circ$)  liquid, ($\blacksquare$) bcc crystal and 
($\blacktriangle$) gels. Dashed line represent equilibrium phase diagram from 
Ref.\cite{likos:prl:98}. The solid line is the RY-ideal MCT glass 
line from Ref.\cite{emanu:prl:03}. Inset: 2-dimensional SANS detector picture 
of $f=63$, $\phi$=0.16 evidencing Bragg reflections.}
\end{figure}
Calculating $\eta$ this way for all $f$, we could then compare 
 experimental and  theoretical phase diagrams, as shown in Fig. \ref{fig:phdiag}. 
 The transition from the liquid to 
the crystalline state (body centered cubic, bcc) predicted by theory
 is perfectly reproduced in experiments for $f=63$ and $f=67$. 
The experimental critical packing fraction for crystallization is found to be $\eta \approx 0.21$. 
Direct evidence for crystal formation comes from 
the presence of Bragg peaks in the 2-dimensional SANS pattern as shown in the inset 
of Fig. \ref{fig:phdiag} for $f=63$ and $\phi=0.16$. 
Moreover crossing $\phi^*$ the first peak of $S(Q)$ assumes a value bigger than 2.8, which is 
the minimum value for a freezing transition according to the 
\textit{Hansen-Verlet} criterion \cite{Hansen} (see Fig \ref{fig:sq-scaling}). 
The peak is further growing with increasing $\phi$.  
Additionally for $\phi>\phi^*$ a third small peak is 
forming, and the  position ratio between the three peaks is 
1:$\sqrt{2}$:$\sqrt{3}$ corresponding to a simple cubic (sc) or bcc lattice. 
The A15 lattice can be directly excluded, since its additional third reflection at $\sqrt{5/2}$ 
is expected to occur in a region where the form factor is close to unity.
A comparison of the lattice constants $a_{sc}=270\AA$ and $a_{bcc}=381\AA$ with $2 R_m=377\AA$ at $\Phi^*$ 
excludes the sc lattice, identifying the crystal as a bcc lattice. 
With increasing $\eta$ no evidence for the predicted transition to a fcc phase is found in our data. 
However, we also note that for regular star polymers a bcc-fcc transition was never observed, 
probably due to the small free energy difference, $\approx kT$, between 
the two crystalline phases \cite{watzlawek:phd}. 

With increasing $f$ the experimentally observed liquid-solid 
transition is in nearly perfect agreement with the phase boundary predicted by theory,  
but instead of observing a fcc phase
 we observe disordered solid-like structures for all $f\ge73$.
Location of the solidus line comes from a double 
evidence: First from tube inversion, which means that we tested 
the non-flowing behavior of the sample inverting sample holders in a 
bath at constant temperature. 
In addition, preliminary rheology experiments confirm a non-zero storage modulus. 
Structure factor peaks for $\phi > \phi^*$ are definitely 
smaller than 2.8, excluding formation of a crystal, as shown in Fig \ref{fig:sq-scaling}. 
Therefore we identify this disordered phase as a glass. 
The suppression of the expected equilibrium fcc phase and the early onset of glass formation 
at packing fractions smaller than expected from theory, might arise 
from an "effective" polydispersity induced by the clusters discussed above \cite{sciortino:prl:2004}. 
At lower $f$, where MCT glass line and crystal line are well separated, 
this effect does not play such an important role and we do observe the crystal there

In conclusion, we have shown that star-like PEP-PEO micelles 
 show the same interactions as star polymers, 
 giving a significant experimental support to the picture of star polymers as ultra-soft colloids. 
We have been able to quantitatively describe experimental structure factors $S(Q)$ starting from a 
microscopic effective potential predicted for star polymers. 
The softness of the interaction potential between star-like micelles can be 
precisely varied by adjusting the interfacial tension $\gamma$ between PEP and the used solvent. 
In addition, we performed a  detailed investigation of the phase diagram as a 
function of functionality $f$ and packing fraction.
In particular we determined the critical packing fraction for liquid-bcc crystal transition, finding 
excellent agreement with theory. 
The simple synthesis of PEP-PEO block copolymers compared to  star polymers in combination with the shown analogy 
in terms of effective interactions establish these micelles an excellent alternative for investigating the phase behaviour of ultra-soft colloids.

We acknowledge the allocation of SANS beam time by FZJ, ILL and LLB and 
assistance during the SANS experiments by A. Radulescu (FZJ), 
P. Lindner (ILL), and  L. Noirez (LLB). 
We acknowledge helpful discussions with C.N. Likos.
This work was supported by the Deutsche Forschungsgemeinschaft in the framework of the 
Transregio-SFB TR6. E. Zaccarelli acknowledges support from MIUR Cofin 2002 and FIRB.
%
\bibliography{laurati_m_prl_2004_revised_2}

\begin{thebibliography}{27}
\expandafter\ifx\csname natexlab\endcsname\relax\def\natexlab#1{#1}\fi
\expandafter\ifx\csname bibnamefont\endcsname\relax
  \def\bibnamefont#1{#1}\fi
\expandafter\ifx\csname bibfnamefont\endcsname\relax
  \def\bibfnamefont#1{#1}\fi
\expandafter\ifx\csname citenamefont\endcsname\relax
  \def\citenamefont#1{#1}\fi
\expandafter\ifx\csname url\endcsname\relax
  \def\url#1{\texttt{#1}}\fi
\expandafter\ifx\csname urlprefix\endcsname\relax\def\urlprefix{URL }\fi
\providecommand{\bibinfo}[2]{#2}
\providecommand{\eprint}[2][]{\url{#2}}

\bibitem[{\citenamefont{Grest et~al.}(1996)\citenamefont{Grest, Fetters, Huang,
  and Richter}}]{grest:acp:96}
\bibinfo{author}{\bibfnamefont{G.}~\bibnamefont{Grest}},
  \bibinfo{author}{\bibfnamefont{L.}~\bibnamefont{Fetters}},
  \bibinfo{author}{\bibfnamefont{J.}~\bibnamefont{Huang}}, \bibnamefont{and}
  \bibinfo{author}{\bibfnamefont{D.}~\bibnamefont{Richter}},
  \bibinfo{journal}{Adv.\ Chem. \ Phys.} \textbf{\bibinfo{volume}{94}},
  \bibinfo{pages}{67} (\bibinfo{year}{1996}).

\bibitem[{\citenamefont{Likos et~al.}(1998)\citenamefont{Likos, L{\"o}wen,
  Watzlawek, Abbas, Jucknische, Allgaier, and D.Richter}}]{likos:prl:98}
\bibinfo{author}{\bibfnamefont{C.}~\bibnamefont{Likos}},
  \bibinfo{author}{\bibfnamefont{H.}~\bibnamefont{L{\"o}wen}},
  \bibinfo{author}{\bibfnamefont{M.}~\bibnamefont{Watzlawek}},
  \bibinfo{author}{\bibfnamefont{B.}~\bibnamefont{Abbas}},
  \bibinfo{author}{\bibfnamefont{O.}~\bibnamefont{Jucknische}},
  \bibinfo{author}{\bibfnamefont{J.}~\bibnamefont{Allgaier}}, \bibnamefont{and}
  \bibinfo{author}{\bibnamefont{D.Richter}}, \bibinfo{journal}{Phys.\ Rev. \
  Lett.} \textbf{\bibinfo{volume}{80}}, \bibinfo{pages}{4450}
  (\bibinfo{year}{1998}).

\bibitem[{\citenamefont{Stellbrink et~al.}(2000)\citenamefont{Stellbrink,
  Allgaier, Monkenbusch, Richter, Lang, Likos, Watzlawek, L{\"o}wen, Ehlers,
  and Schleger}}]{stellbrink:pcps:00}
\bibinfo{author}{\bibfnamefont{J.}~\bibnamefont{Stellbrink}},
  \bibinfo{author}{\bibfnamefont{J.}~\bibnamefont{Allgaier}},
  \bibinfo{author}{\bibfnamefont{M.}~\bibnamefont{Monkenbusch}},
  \bibinfo{author}{\bibfnamefont{D.}~\bibnamefont{Richter}},
  \bibinfo{author}{\bibfnamefont{A.}~\bibnamefont{Lang}},
  \bibinfo{author}{\bibfnamefont{C.}~\bibnamefont{Likos}},
  \bibinfo{author}{\bibfnamefont{M.}~\bibnamefont{Watzlawek}},
  \bibinfo{author}{\bibfnamefont{H.}~\bibnamefont{L{\"o}wen}},
  \bibinfo{author}{\bibfnamefont{G.}~\bibnamefont{Ehlers}}, \bibnamefont{and}
  \bibinfo{author}{\bibfnamefont{P.}~\bibnamefont{Schleger}},
  \bibinfo{journal}{Prog.\ Colloid \ Polym. \ Sci.}
  \textbf{\bibinfo{volume}{115}}, \bibinfo{pages}{88} (\bibinfo{year}{2000}).

\bibitem[{\citenamefont{Watzlawek et~al.}(1999)\citenamefont{Watzlawek, Likos,
  and L{\"o}wen}}]{wat:prl:99}
\bibinfo{author}{\bibfnamefont{M.}~\bibnamefont{Watzlawek}},
  \bibinfo{author}{\bibfnamefont{C.}~\bibnamefont{Likos}}, \bibnamefont{and}
  \bibinfo{author}{\bibfnamefont{H.}~\bibnamefont{L{\"o}wen}},
  \bibinfo{journal}{Phys.\ Rev.\ Lett.} \textbf{\bibinfo{volume}{82}},
  \bibinfo{pages}{5289} (\bibinfo{year}{1999}).

\bibitem[{\citenamefont{Foffi et~al.}(2003)\citenamefont{Foffi, Sciortino,
  Tartaglia, Zaccarelli, Verso, Reatto, Dawson, and Likos}}]{emanu:prl:03}
\bibinfo{author}{\bibfnamefont{G.}~\bibnamefont{Foffi}},
  \bibinfo{author}{\bibfnamefont{F.}~\bibnamefont{Sciortino}},
  \bibinfo{author}{\bibfnamefont{P.}~\bibnamefont{Tartaglia}},
  \bibinfo{author}{\bibfnamefont{E.}~\bibnamefont{Zaccarelli}},
  \bibinfo{author}{\bibfnamefont{F.~L.} \bibnamefont{Verso}},
  \bibinfo{author}{\bibfnamefont{L.}~\bibnamefont{Reatto}},
  \bibinfo{author}{\bibfnamefont{K.}~\bibnamefont{Dawson}}, \bibnamefont{and}
  \bibinfo{author}{\bibfnamefont{C.}~\bibnamefont{Likos}},
  \bibinfo{journal}{Phys.\ Rev.\ Lett.} \textbf{\bibinfo{volume}{90}},
  \bibinfo{pages}{5289} (\bibinfo{year}{2003}).

\bibitem[{\citenamefont{Vlassopoulos et~al.}(2001)\citenamefont{Vlassopoulos,
  Fytas, Pakula, and Roovers}}]{vlass:jpcm:01}
\bibinfo{author}{\bibfnamefont{D.}~\bibnamefont{Vlassopoulos}},
  \bibinfo{author}{\bibfnamefont{G.}~\bibnamefont{Fytas}},
  \bibinfo{author}{\bibfnamefont{T.}~\bibnamefont{Pakula}}, \bibnamefont{and}
  \bibinfo{author}{\bibfnamefont{J.}~\bibnamefont{Roovers}},
  \bibinfo{journal}{J. Phys: Condens. Matter} \textbf{\bibinfo{volume}{13}},
  \bibinfo{pages}{R855} (\bibinfo{year}{2001}).

\bibitem[{\citenamefont{Kapnistos et~al.}(2000)\citenamefont{Kapnistos,
  Vlassopoulos, Fytas, Mortensen, Fleischer, and Roovers}}]{kapnistos2001}
\bibinfo{author}{\bibfnamefont{M.}~\bibnamefont{Kapnistos}},
  \bibinfo{author}{\bibfnamefont{D.}~\bibnamefont{Vlassopoulos}},
  \bibinfo{author}{\bibfnamefont{G.}~\bibnamefont{Fytas}},
  \bibinfo{author}{\bibfnamefont{K.}~\bibnamefont{Mortensen}},
  \bibinfo{author}{\bibfnamefont{G.}~\bibnamefont{Fleischer}},
  \bibnamefont{and} \bibinfo{author}{\bibfnamefont{J.}~\bibnamefont{Roovers}},
  \bibinfo{journal}{Phys. \ Rev. \ Lett.} \textbf{\bibinfo{volume}{85}},
  \bibinfo{pages}{4072} (\bibinfo{year}{2000}).

\bibitem[{\citenamefont{Hamley}(1998)}]{hamley:1998}
\bibinfo{author}{\bibfnamefont{I.}~\bibnamefont{Hamley}},
  \emph{\bibinfo{title}{The Physics of Block Copolymers}}
  (\bibinfo{publisher}{Oxford University Press}, \bibinfo{year}{1998}).

\bibitem[{\citenamefont{Halperin et~al.}(1992)\citenamefont{Halperin, Tirell,
  and Lodge}}]{halperin:aps:1992}
\bibinfo{author}{\bibfnamefont{A.}~\bibnamefont{Halperin}},
  \bibinfo{author}{\bibfnamefont{M.}~\bibnamefont{Tirell}}, \bibnamefont{and}
  \bibinfo{author}{\bibfnamefont{T.}~\bibnamefont{Lodge}},
  \bibinfo{journal}{Adv. Polym. Sci.} \textbf{\bibinfo{volume}{100}},
  \bibinfo{pages}{31} (\bibinfo{year}{1992}).

\bibitem[{\citenamefont{McConnell et~al.}(1993)\citenamefont{McConnell, Gast,
  Huang, and Smith}}]{gast:prl:93}
\bibinfo{author}{\bibfnamefont{G.}~\bibnamefont{McConnell}},
  \bibinfo{author}{\bibfnamefont{A.}~\bibnamefont{Gast}},
  \bibinfo{author}{\bibfnamefont{J.}~\bibnamefont{Huang}}, \bibnamefont{and}
  \bibinfo{author}{\bibfnamefont{S.}~\bibnamefont{Smith}},
  \bibinfo{journal}{Phys. \ Rev. \ Lett.} \textbf{\bibinfo{volume}{71}},
  \bibinfo{pages}{2102} (\bibinfo{year}{1993}).

\bibitem[{\citenamefont{Lodge et~al.}(2004)\citenamefont{Lodge, Bang, Park, and
  Char}}]{lodge:prl:04}
\bibinfo{author}{\bibfnamefont{T.}~\bibnamefont{Lodge}},
  \bibinfo{author}{\bibfnamefont{J.}~\bibnamefont{Bang}},
  \bibinfo{author}{\bibfnamefont{M.}~\bibnamefont{Park}}, \bibnamefont{and}
  \bibinfo{author}{\bibfnamefont{K.}~\bibnamefont{Char}},
  \bibinfo{journal}{Phys. \ Rev. \ Lett.} \textbf{\bibinfo{volume}{92}},
  \bibinfo{pages}{145501} (\bibinfo{year}{2004}).

\bibitem[{\citenamefont{Hamley et~al.}(1998)\citenamefont{Hamley, Pople, and
  Diat}}]{hamley:cps:98}
\bibinfo{author}{\bibfnamefont{I.}~\bibnamefont{Hamley}},
  \bibinfo{author}{\bibfnamefont{J.}~\bibnamefont{Pople}}, \bibnamefont{and}
  \bibinfo{author}{\bibfnamefont{O.}~\bibnamefont{Diat}},
  \bibinfo{journal}{Colloid Polym. Sci.} \textbf{\bibinfo{volume}{276}},
  \bibinfo{pages}{446} (\bibinfo{year}{1998}).

\bibitem[{\citenamefont{Beaudoin et~al.}(2002)\citenamefont{Beaudoin, Borisov,
  Lapp, Billon, Hiorns, and Francois}}]{beaudoin02}
\bibinfo{author}{\bibfnamefont{E.}~\bibnamefont{Beaudoin}},
  \bibinfo{author}{\bibfnamefont{O.}~\bibnamefont{Borisov}},
  \bibinfo{author}{\bibfnamefont{A.}~\bibnamefont{Lapp}},
  \bibinfo{author}{\bibfnamefont{L.}~\bibnamefont{Billon}},
  \bibinfo{author}{\bibfnamefont{R.}~\bibnamefont{Hiorns}}, \bibnamefont{and}
  \bibinfo{author}{\bibfnamefont{J.}~\bibnamefont{Francois}},
  \bibinfo{journal}{Macromolecules} \textbf{\bibinfo{volume}{35}},
  \bibinfo{pages}{7436} (\bibinfo{year}{2002}).

\bibitem[{\citenamefont{Willner et~al.}(2000)\citenamefont{Willner, Poppe,
  Allgaier, Monkenbusch, Lindner, and Richter}}]{willner:epl:00}
\bibinfo{author}{\bibfnamefont{L.}~\bibnamefont{Willner}},
  \bibinfo{author}{\bibfnamefont{A.}~\bibnamefont{Poppe}},
  \bibinfo{author}{\bibfnamefont{J.}~\bibnamefont{Allgaier}},
  \bibinfo{author}{\bibfnamefont{J.}~\bibnamefont{Monkenbusch}},
  \bibinfo{author}{\bibfnamefont{P.}~\bibnamefont{Lindner}}, \bibnamefont{and}
  \bibinfo{author}{\bibfnamefont{D.}~\bibnamefont{Richter}},
  \bibinfo{journal}{Europhys. Lett.} \textbf{\bibinfo{volume}{51}},
  \bibinfo{pages}{628} (\bibinfo{year}{2000}).

\bibitem[{\citenamefont{Lund et~al.}(2004)\citenamefont{Lund, Willner,
  Stellbrink, Radulescu, and Richter}}]{reidar:macro}
\bibinfo{author}{\bibfnamefont{R.}~\bibnamefont{Lund}},
  \bibinfo{author}{\bibfnamefont{L.}~\bibnamefont{Willner}},
  \bibinfo{author}{\bibfnamefont{J.}~\bibnamefont{Stellbrink}},
  \bibinfo{author}{\bibfnamefont{A.}~\bibnamefont{Radulescu}},
  \bibnamefont{and} \bibinfo{author}{\bibfnamefont{D.}~\bibnamefont{Richter}},
  \bibinfo{journal}{Macromolecules} \textbf{\bibinfo{volume}{37}},
  \bibinfo{pages}{9984} (\bibinfo{year}{2004}).

\bibitem[{\citenamefont{Stellbrink et~al.}(2004)\citenamefont{Stellbrink,
  Rother, Laurati, Lund, L.Willner, and Richter}}]{stellbrink:jpcm}
\bibinfo{author}{\bibfnamefont{J.}~\bibnamefont{Stellbrink}},
  \bibinfo{author}{\bibfnamefont{G.}~\bibnamefont{Rother}},
  \bibinfo{author}{\bibfnamefont{M.}~\bibnamefont{Laurati}},
  \bibinfo{author}{\bibfnamefont{R.}~\bibnamefont{Lund}},
  \bibinfo{author}{\bibnamefont{L.Willner}}, \bibnamefont{and}
  \bibinfo{author}{\bibfnamefont{D.}~\bibnamefont{Richter}},
  \bibinfo{journal}{J. Phys.: Condens. Matter} \textbf{\bibinfo{volume}{16}},
  \bibinfo{pages}{S3821} (\bibinfo{year}{2004}).

\bibitem[{\citenamefont{Poppe et~al.}(1997)\citenamefont{Poppe, Willner,
  Allgaier, Stellbrink, and Richter}}]{poppe:macro:97}
\bibinfo{author}{\bibfnamefont{A.}~\bibnamefont{Poppe}},
  \bibinfo{author}{\bibfnamefont{L.}~\bibnamefont{Willner}},
  \bibinfo{author}{\bibfnamefont{J.}~\bibnamefont{Allgaier}},
  \bibinfo{author}{\bibfnamefont{J.}~\bibnamefont{Stellbrink}},
  \bibnamefont{and} \bibinfo{author}{\bibfnamefont{D.}~\bibnamefont{Richter}},
  \bibinfo{journal}{Macromolecules} \textbf{\bibinfo{volume}{30}},
  \bibinfo{pages}{7462} (\bibinfo{year}{1997}).

\bibitem[{\citenamefont{Lund}(2004)}]{reidar:thesis}
\bibinfo{author}{\bibfnamefont{R.}~\bibnamefont{Lund}}, Ph.D. thesis,
  \bibinfo{school}{Universit\"at M\"unster} (\bibinfo{year}{2004}).

\bibitem[{\citenamefont{Stiakakis et~al.}(2002)\citenamefont{Stiakakis,
  Vlassopoulos, Likos, Roovers, and Meier}}]{stiakakis:prl:02}
\bibinfo{author}{\bibfnamefont{E.}~\bibnamefont{Stiakakis}},
  \bibinfo{author}{\bibfnamefont{D.}~\bibnamefont{Vlassopoulos}},
  \bibinfo{author}{\bibfnamefont{C.}~\bibnamefont{Likos}},
  \bibinfo{author}{\bibfnamefont{J.}~\bibnamefont{Roovers}}, \bibnamefont{and}
  \bibinfo{author}{\bibfnamefont{G.}~\bibnamefont{Meier}},
  \bibinfo{journal}{Phys. \ Rev. \ Lett.} \textbf{\bibinfo{volume}{89}},
  \bibinfo{pages}{208302} (\bibinfo{year}{2002}).

\bibitem[{\citenamefont{Stellbrink et~al.}(1997)\citenamefont{Stellbrink,
  Allgaier, and Richter}}]{stellbrink97}
\bibinfo{author}{\bibfnamefont{J.}~\bibnamefont{Stellbrink}},
  \bibinfo{author}{\bibfnamefont{J.}~\bibnamefont{Allgaier}}, \bibnamefont{and}
  \bibinfo{author}{\bibfnamefont{D.}~\bibnamefont{Richter}},
  \bibinfo{journal}{Phys.\ Rev. \ E} \textbf{\bibinfo{volume}{56}},
  \bibinfo{pages}{R3772} (\bibinfo{year}{1997}).

\bibitem[{\citenamefont{Verso et~al.}(2003)\citenamefont{Verso, Tau, and
  Reatto}}]{federica:jpcm:03}
\bibinfo{author}{\bibfnamefont{F.~L.} \bibnamefont{Verso}},
  \bibinfo{author}{\bibfnamefont{M.}~\bibnamefont{Tau}}, \bibnamefont{and}
  \bibinfo{author}{\bibfnamefont{L.}~\bibnamefont{Reatto}},
  \bibinfo{journal}{J.Phys: Condens. Matter} \textbf{\bibinfo{volume}{15}},
  \bibinfo{pages}{1505} (\bibinfo{year}{2003}).

\bibitem[{\citenamefont{Watzlawek et~al.}(1998)\citenamefont{Watzlawek,
  L{\"o}wen, and Likos}}]{watz:jpcm:98}
\bibinfo{author}{\bibfnamefont{M.}~\bibnamefont{Watzlawek}},
  \bibinfo{author}{\bibfnamefont{H.}~\bibnamefont{L{\"o}wen}},
  \bibnamefont{and} \bibinfo{author}{\bibfnamefont{C.}~\bibnamefont{Likos}},
  \bibinfo{journal}{J.Phys: Condens. Matter} \textbf{\bibinfo{volume}{10}},
  \bibinfo{pages}{8189} (\bibinfo{year}{1998}).

\bibitem[{\citenamefont{Pedersen et~al.}(1990)\citenamefont{Pedersen, Posselt,
  and Mortensen}}]{pedersen}
\bibinfo{author}{\bibfnamefont{J.}~\bibnamefont{Pedersen}},
  \bibinfo{author}{\bibfnamefont{D.}~\bibnamefont{Posselt}}, \bibnamefont{and}
  \bibinfo{author}{\bibfnamefont{K.}~\bibnamefont{Mortensen}},
  \bibinfo{journal}{J. Appl. Cryst.} \textbf{\bibinfo{volume}{23}},
  \bibinfo{pages}{321} (\bibinfo{year}{1990}).

\bibitem[{\citenamefont{Daoud and Cotton}(1982)}]{daoud-cotton}
\bibinfo{author}{\bibfnamefont{M.}~\bibnamefont{Daoud}} \bibnamefont{and}
  \bibinfo{author}{\bibfnamefont{J.}~\bibnamefont{Cotton}},
  \bibinfo{journal}{J. Phys. (Paris)} \textbf{\bibinfo{volume}{43}},
  \bibinfo{pages}{531} (\bibinfo{year}{1982}).

\bibitem[{\citenamefont{Hansen and Verlet}(1969)}]{Hansen}
\bibinfo{author}{\bibfnamefont{J.}~\bibnamefont{Hansen}} \bibnamefont{and}
  \bibinfo{author}{\bibfnamefont{L.}~\bibnamefont{Verlet}},
  \bibinfo{journal}{Phys. Rev.} \textbf{\bibinfo{volume}{184}},
  \bibinfo{pages}{151} (\bibinfo{year}{1969}).

\bibitem[{\citenamefont{Watzlawek}(1999)}]{watzlawek:phd}
\bibinfo{author}{\bibfnamefont{M.}~\bibnamefont{Watzlawek}}, Ph.D. thesis,
  \bibinfo{school}{University of Dusseldorf} (\bibinfo{year}{1999}).

\bibitem[{\citenamefont{Sciortino et~al.}(2004)\citenamefont{Sciortino, Mossa,
  Zaccarelli, and Tartaglia}}]{sciortino:prl:2004}
\bibinfo{author}{\bibfnamefont{F.}~\bibnamefont{Sciortino}},
  \bibinfo{author}{\bibfnamefont{S.}~\bibnamefont{Mossa}},
  \bibinfo{author}{\bibfnamefont{E.}~\bibnamefont{Zaccarelli}},
  \bibnamefont{and}
  \bibinfo{author}{\bibfnamefont{P.}~\bibnamefont{Tartaglia}},
  \bibinfo{journal}{Phys. \ Rev. \ Lett.} \textbf{\bibinfo{volume}{93}},
  \bibinfo{pages}{055701} (\bibinfo{year}{2004}).

\end{thebibliography}
\end{document}